\documentclass[10pt,conference]{IEEEtran}
\usepackage[utf8]{inputenc}
\usepackage{cite}
\usepackage{soul}
\usepackage{xspace}
\usepackage{url}
\usepackage{graphicx}
\usepackage{listings,multicol}
\usepackage[caption=false]{subfig}
\usepackage[export]{adjustbox}
\usepackage{amsmath}
\usepackage{textcomp}
\usepackage{breakurl} 
\usepackage[dvipsnames]{xcolor}
\usepackage{listings}
\usepackage{etoolbox}
\usepackage{hyperref}
\usepackage{cleveref}

\lstdefinelanguage{Kotlin}{
  comment=[l]{//},
  commentstyle={\color{gray}\ttfamily},
  emph={delegate, filter, first, firstOrNull, forEach, lazy, map, mapNotNull, println, return@},
  emphstyle={\color{OrangeRed}},
  identifierstyle=\color{black},
  keywords={abstract, actual, as, as?, break, by, class, companion, continue, data, do, dynamic, else, enum, expect, false, final, for, fun, get, if, import, in, interface, internal, is, null, object, override, package, private, public, return, set, super, suspend, this, throw, true, try, typealias, val, var, vararg, when, where, while},
  keywordstyle={\color{NavyBlue}\bfseries},
  morecomment=[s]{/*}{*/},
  morestring=[b]",
  morestring=[s]{"""*}{*"""},
  ndkeywords={@Deprecated, @JvmField, @JvmName, @JvmOverloads, @JvmStatic, @JvmSynthetic, Array, Byte, Double, Float, Int, Integer, Iterable, Long, Runnable, Short, String},
  ndkeywordstyle={\color{BurntOrange}\bfseries},
  sensitive=true,
  stringstyle={\color{ForestGreen}\ttfamily},
}

\providetoggle{draft}

\newcommand{\todo}[1]{\iftoggle{draft}{\hl{\textbf{TODO: ``#1''}}\xspace}{}}

\newcommand{\eg}{\emph{e.g.,}\xspace}
\newcommand{\miner}{TNM\xspace}

\title{TNM: A Tool for Mining of Socio-Technical Data from Git Repositories}
\author{\IEEEauthorblockN{Nikolai Sviridov}
\IEEEauthorblockA{\textit{Department of IT and Programming} \\
\textit{ITMO University}\\
St. Petersburg, Russia \\
nikolaisvg@gmail.com}
\and
\IEEEauthorblockN{Mikhail Evtikhiev}
\IEEEauthorblockA{\textit{Intelligent Collaboration Tools Lab} \\
\textit{JetBrains Research}\\
St. Petersburg, Russia \\
Mikhail.Evtikhiev@jetbrains.com}
\and
\IEEEauthorblockN{Vladimir Kovalenko}
\IEEEauthorblockA{\textit{Intelligent Collaboration Tools Lab} \\
\textit{JetBrains Research}\\
Amsterdam, The Netherlands \\
Vladimir.Kovalenko@jetbrains.com}
}

\begin{document}
\maketitle

\begin{abstract}
Networks of collaboration between engineers 
are reflected in traces of developers' activity in version control systems (VCSs). 
Extracting data from Git repositories is an essential task for researchers and practitioners working on socio-technical analysis, but it requires substantial engineering work. 
With increasing interest in analysing socio-technical data and applying it in practice, there are no flexible and easily reusable tools to retrieve socio-technical information from VCSs. 
With no common reusable toolkit existing for this task, the burden of mining diverts the focus of researchers from their core research questions.

In this paper, we present TNM---an open-source tool for mining socio-technical data from Git repositories. TNM is fast, flexible, and easily extensible.

\textbf{\miner is available on GitHub: \url{https://github.com/JetBrains-Research/tnm}}
\end{abstract}


\section{Introduction}

Many software projects are products of teamwork. 
According to the ISBSG repository~\cite{isbsg}, the average size of a development team, averaged over time, is 7.9 members and the median size is 5~\cite{rodriguez2012empirical}. 
While teamwork is often a boon that allows developers to have fruitful discussions and create better products~\cite{rahman2011ownership}, it also requires coordination and communication between team members. 
Poor communication can result in inconsistent design solutions and bugs and generally impede the development of the project~\cite{stelzer1998success}.
Thus, for teamwork to be successful, it is important for managers and peers to have an understanding of communication and coordination in the team and existing issues with interactions between teammates and understand how these issues can be addressed~\cite{hoegl2001teamwork}.

There is a large and growing body of research dedicated to studying interactions between developers.
Approaches undertaken in these studies can be broadly divided into two groups. 
First, it is possible to carry out a study of a group 
of engineers to understand their interactions and deduce some patterns leading to a better understanding of collaborative work.
An alternative way is to collect data on interaction of developers and analyze it to devise automatic processing approaches that can be used to produce actionable insights~\cite{cosentino}.


One major source of socio-technical data is version control repositories (VCS repositories). 
VCSs such as Git~\cite{git} store information about the state of the system after each commit, along with the information about the developer who made the commit.
It is possible to recover from project VCS repositories~\cite{gall2003cvs} the structure of the project and logical connections between its parts such as
logical coupling~\cite{gall1998detection}. 
Furthermore, the structure of projects often mirrors the collaboration network of its contributors~\cite{herbsleb1999architectures, Conway1967HOWDC}). 
Thus, information from VCSs can be used as a proxy to derive interactions between the developers at least to an extent.

A number of prior studies investigated collaboration between engineers using information extracted from VCS repositories. 
Examples include computing the degree of knowledge to study developers' contributions~\cite{carlson2015engaging, avelino2019measuring}, analyzing code ownership to identify bug-prone components~\cite{rahman2011ownership, meneely2009secure}, estimating existential risks of projects~\cite{cosentino, avelino2016novel, rigby2016quantifying, yamashita2015revisiting}, and applying the PageRank~\cite{googlePageRank} algorithm to infer causal relationships between the commits~\cite{pagerank-commits}. 

Despite the fact that socio-technical data is a popular area of study, there is a lack of publicly available tools to mine such data, so researchers have to spend time and effort to develop their own.
Existing open-source tools for data extraction from Git repositories (\eg \cite{RepoDriller}, \cite{PyDriller}, and \cite{duenas2018perceval}) allow to mine mostly raw (non-processed) data, which requires extra processing.
Thus, the lack of common toolkits for extraction of and manipulation with socio-technical data constitutes a hurdle that diverts the focus of researchers from their core research questions.

With this observation in mind, we created \miner---an open-source tool for mining socio-technical data from Git repositories.
\miner is a fast tool written in Kotlin~\cite{kotlin} and can be used as an external library. 
\miner works with local Git repositories. 
\miner incorporates implementations of several established data mining techniques, or individual \emph{miners}. 
Namely, it includes a miner for the degree of knowledge-based files ownership~\cite{carlson2015engaging}, a miner applying the PageRank algorithm~\cite{googlePageRank} to the analysis of a commit history for commit influence evaluation~\cite{pagerank-commits}, miners extracting data for the analysis of socio-technical congruence~\cite{congruence}, a miner for calculating the distribution of commit time over the week, and one for finding files that have been edited by several developers.
Moreover, \miner can be easily extended to include other analysis algorithms.

The primary contribution of this work is \miner---a tool that can be used to mine socio-technical data as a standalone tool or integrated into an existing mining pipeline. 

\section{Background and Motivation}
\subsection{Socio-Technical Data}
An average software user wants their products to be bug-free\cite{simmonds2018complexity}. 
However, due to the complex nature of software systems, it is hard to develop bug-free software~\cite{simmonds2018complexity}. 
Some development patterns have been shown to correlate with an increased risk of introducing new bugs~\cite{pagerank-commits, rahman2011ownership, fritz2010degree, meneely2009secure} and increased existential risks of the project~\cite{avelino2016novel, cosentino, rigby2016quantifying}.

A particular kind of data that can be used to identify possible sources of project problems is socio-technical data closely aligned with the framework of socio-technical congruence~\cite{cataldo2008socio, congruence}. 
The core assumption of socio-technical congruence is that coordination activities of the developers should be in agreement with the coordination requirements, and the coordination requirements are determined by the project development. 
Some approaches to identifying problematic parts of software projects are implicitly using socio-technical congruence, see, \eg~\cite{avelino2019measuring, fritz2010degree, yamashita2015revisiting}.

For our purposes, we define socio-technical data as a combination of data that reflects interactions between developers (the ``socio-'' part), and data that corresponds to the programming activities performed by the engineers and the technical dependencies between various components of the project (the ``technical'' part)\cite{cataldo2008socio}. 
Socio-technical data includes, in particular, information on project evolution and, therefore, may be leveraged to identify the origins of problems that arise during  project development.

\subsection{Version Control Systems as Data Sources}
VCS repositories store traces of most development and maintenance efforts in software projects and, therefore, are an indispensable source of socio-technical data~\cite{bird2009promises}, allowing researchers to collect rich datasets for their studies. 

Git~\cite{git} is by far the most popular VCS today.
A core concept in Git's data model is a commit---a snapshot of a repository reflecting the state of its file tree.
In addition to the state of the files, commits contain metadata such as the author, timestamp, and links to parent commits representing one or several previous states.
This metadata is particularly important for socio-technical analysis, since it allows to trace individual contributions of developers and the proximity of their contributions by going through the history.


\subsection{Analysis Techniques}
Methods to infer socio-technical data from Git repositories include applying the PageRank algorithm to evaluate the influence of a commit on other commits~\cite{pagerank-commits}, computing the degree of knowledge to evaluate the properties of developers' contributions~\cite{avelino2019measuring, carlson2015engaging}, analyzing collaboration in a project to evaluate its risk of stalling~\cite{cosentino, avelino2016novel, rigby2016quantifying, yamashita2015revisiting}, determining code ownership and code authorship to find code components and source files at risk of having quality issues~\cite{rahman2011ownership, meneely2009secure}, and using socio-technical congruence to measure the alignment between the coordination required by the project technical dependencies and the actual coordination between its members~\cite{congruence, conway}. 

\subsubsection{Degree of Knowledge and Degree of Interest}\label{sec:dok-doi}
Evaluating the properties of developers' contributions to projects and the distribution of developers' knowledge is a long-standing problem~\cite{mcdonald2000expertise, mockus2002expertise} with many possible applications from familiarizing new team members with relevant parts of the project 
to identifying interesting bug reports~\cite{fritz2010degree}. 
\emph{Degree of knowledge} is a metric first introduced by Fritz et al.~\cite{fritz2010degree}, that consists of two parts.
\emph{Degree of authorship} part is a measure of developers' contribution to code, that can be computed with the data from git repository.
\emph{Degree of interest} part is a measure of developers' interactions with code; it is necessary to track developers' behavior to calculate it. 
Some of the later implementations of the degree of knowledge  rely only on data from Git~\cite{avelino2019measuring, carlson2015engaging}. 
The obtained degree of knowledge can be then used to study individual contribution patterns
~\cite{avelino2019measuring} or to compute the bus factor of the project~\cite{avelino2016novel} (see \Cref{sec:bus-factor}).

\subsubsection{PageRank}
PageRank algorithm was originally developed by Google~\cite{googlePageRank} to evaluate web pages popularity. 
The more web pages refer to a given web page, the higher score it gets, and the web pages that are referred to by a~web page with a high score also get high scores. 
PageRank can be applied to arbitrary linked entities. 
Suzuki et al.~\cite{pagerank-commits} have applied it to commits to infer causal relations between the commits and evaluate how commits influence other successive commits. 
The researchers focused on bug-introducing and bug-fixing commits to study what kinds of commits are more bug-prone.

\subsubsection{Code Ownership}
Code ownership reflects the
distribution of code authorship between developers at various levels of granularity.
Git repositories are a major source of authorship data~\cite{meneely2009secure, rahman2011ownership}. 
Studies of code ownership have unearthed important patterns across many projects: \eg higher code ownership 
is correlated with higher component quality~\cite{bird2011don}; 
code changed in a bugfix is less likely to have contributions from multiple authors, and better experience with a certain file makes a developer less likely to introduce defects into it~\cite{rahman2011ownership}.

\subsubsection{Bus Factor}\label{sec:bus-factor}
Bus factor, also known as truck factor,
is defined as the minimal number of developers whose leaving the project (being ``hit by a bus'') would cause the project to stall.
While the common definition of the bus factor is hard to quantify, it is possible to create metrics computed from projects' VCS history that closely match developers' perception of the bus factor~\cite{ferreira2019algorithms}.
Such metrics are usually based on utilizing information about code authorship computed at a line or a file level~\cite{cosentino, avelino2016novel, rigby2016quantifying} and may make use of the degree of authorship concept~\cite{avelino2016novel} (\Cref{sec:dok-doi}).
The risk of stalling is calculated based on the
amount of abandoned code.
This simple risk-gauging approach can be further enhanced to suggest potential successors to own abandoned code~\cite{rigby2016quantifying}, enabling project members to mitigate existential risks. 
It is also possible to expand this approach to estimate knowledge loss and persistence of abandoned files~\cite{nassif2017revisiting}.

\subsubsection{Socio-Technical Congruence}\label{sec:congruence-description}
    Socio-technical congruence can be construed 
    as a measure of similarity between the network of technical dependencies within a project and the structure of communication in its team~\cite{congruence}.
    One way to define such metric is through the concept of a coordination need.
    A \emph{coordination need} indicates that two people should be coordinating based on the technical dependencies and is determined by analyzing the assignments of people to a technical entity such as a source code module and the technical dependencies across such technical entities~\cite{congruence}. 
    
\begin{figure}[htp]
        \centering
        \includegraphics[width=4cm]{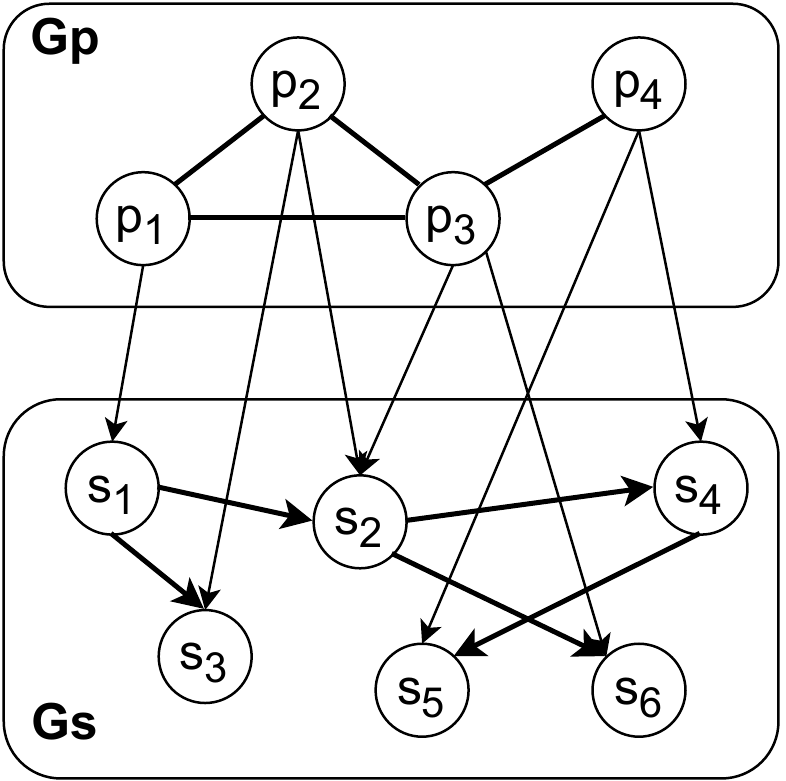}
        \caption{A socio-technical network. Data collected from Git repositories can be used to build the $G_s$ part of the graph and the edges between $G_p$ and $G_s$} \label{fig:stsn}.
    \end{figure}

    If two engineers have a coordination need but do not coordinate,
    a coordination gap may exist, and one of the goals of the socio-technical congruence analysis is to reduce the number of gaps by either aiding coordination 
    or reducing the number of coordination needs~\cite{sarma2008challenges}.
    In the approach of Valetto et al.~\cite{conway}, the network for calculation of congruence unites three different classes of information~(\Cref{fig:stsn}):
    \begin{itemize}
        \item Information about communication between project members 
        represented by directed edges on the $G_p$ (\emph{people}) layer.
        \item Information about relationships between technical entities,
        represented by directed edges on the $G_s$ (\emph{software}) layer.
        \item Information about the contributions of participants to the project entities, represented by directed edges from nodes on $G_p$ plane to nodes on $G_s$ plane.
    \end{itemize}
   The resulting socio-technical network can be used to calculate congruence as the degree of alignment between social relationships and software relationships.     

\subsection{Motivation}
Socio-technical data analysis is a field that naturally attracts attention of researchers from other disciplines.
Moreover, one can argue that the hypotheses proven true for the software developers community may also hold to a certain extent for other communities of practice. 
For example, the socio-technical congruence framework ~\cite{cataldo2008socio} is a development of a study by Conway~\cite{Conway1967HOWDC}.
However, tools to mine data from VCS repositories are not easy to obtain: researchers do not always share their tools, or the prototypes may be hard to adapt in new contexts. The available open-source tools such as PyDriller~\cite{PyDriller} or Perceval~\cite{duenas2018perceval} are not directly aimed at extracting socio-technical data.
Implementing one's own data extraction and processing tools is time-consuming and can be especially complicated for a researcher from a community where mining software repositories is not a universal skill.
This significant technical work required for data mining diverts the focus of researchers and developers from their domain.


This observation motivated us to create \miner---a flexible and easily extendable tool for mining socio-technical data from Git repositories that allows researchers and practitioners to avoid implementing complex mining pipelines and, instead, focus on their problem domain.

\section{Overview of \miner}

The purpose of \miner is to mine several types of socio-technical data from Git repositories. The value of the tool is in the reduction of development effort for extracting socio-technical data for further analysis.
In this section, we describe the inner workings of \miner in more detail.

\subsection{Technologies in Use}

\miner is written in Kotlin.
It can be used either as a standalone tool or as an external library for JVM applications. 
\miner works with local repositories through JGit~\cite{jgit}.
For complex calculations such as matrix multiplication, \miner relies on DL4J~\cite{dl4j}. 
\miner features basic visualization capabilities based on \emph{vis-network}~\cite{visj}.


\subsection{Internal Structure}

\begin{figure}[htp]
    \centering
    \includegraphics[width=.9\columnwidth]{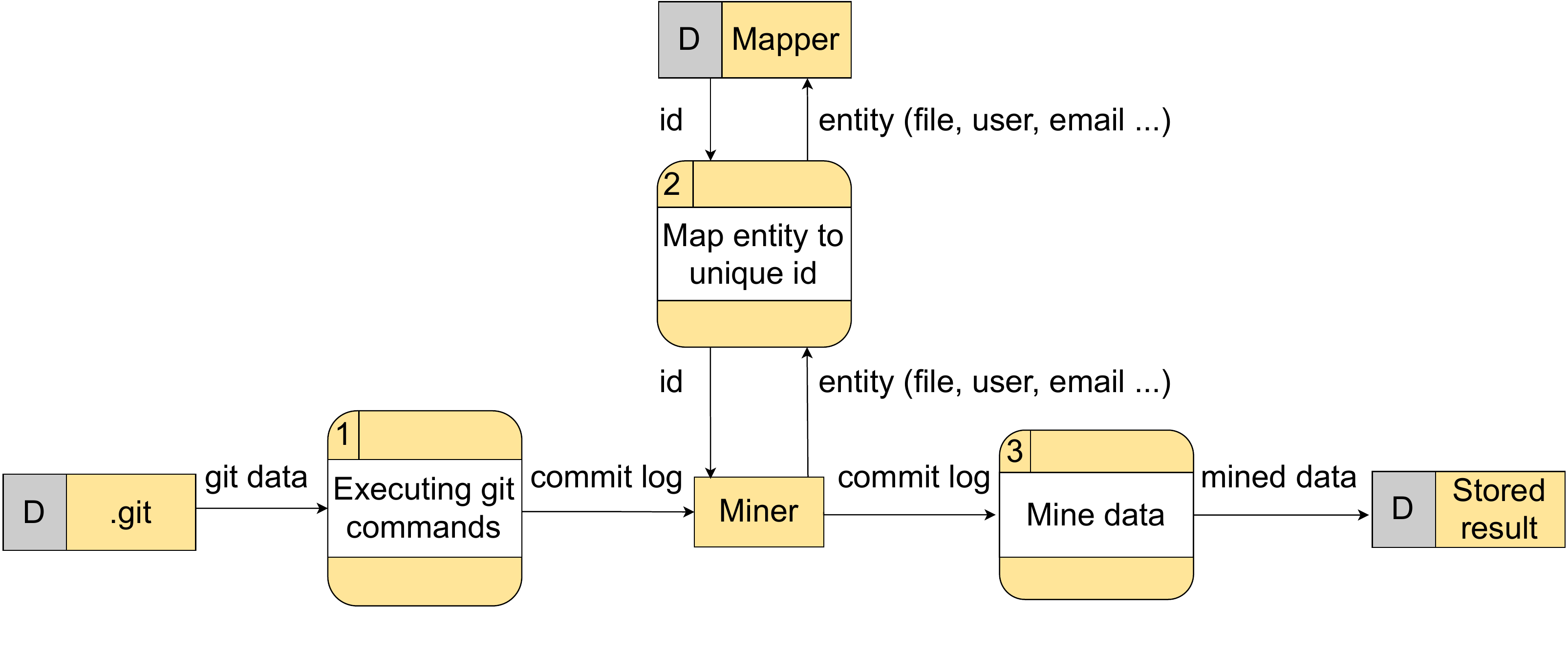}
    \caption{Overview of \miner data flow. \todo{increase font size to be readable}} 
\end{figure}

The main idea behind the structure of \miner is that 
each task is performed by a dedicated class implementing a common interface. 
This allows to reuse code responsible for frequent operations such as traversal of Git commit trees.

\textbf{Mappers}. Mappers are singleton classes that map each data entity to a unique numerical identifier. 
Using numerical identifiers allows to easily render data in a matrix form,
with each column or row representing the data and the value in a specific cell representing a parameter efficiently capturing relationships between pairs of entities. 
All mappers implement the \textit{Mapper} interface. 
Currently, three different mappers are implemented in \miner:
\begin{itemize}
    \item \textit{CommitMapper} maps an SHA-1 commit identifier to a unique numerical value.
    \item \textit{FileMapper} maps a filepath to a unique numerical value.
    \item \textit{UserMapper} maps a developer identity to a unique numerical value.
\end{itemize}
The \textit{Mapper} interface defines two main functions, \textit{add} and \textit{saveToJson}.
The \textit{add} function adds an entity of data to the map and returns its unique numeric ID. \textit{saveToJson} saves the map to JSON.

\textbf{Git Miners}. Git miners are classes implementing the mining tasks. 
All miners use a local Git repository to extract data,
and extend the abstract class \textit{GitMiner} with two functions to process commit history in chosen branches and save the~results. 

The current version of \miner includes six GitMiner implementations that we describe below.

\begin{itemize}
    \item \textit{FilesOwnershipMiner} is based on the Degree of Knowledge (DOK)~\cite{carlson2015engaging}, 
    which measures the knowledge of a developer or set of developers about a particular section of code.
    The miner yields a knowledge score for every \emph{developer + file} pair
    in the form of a nested map, and information about the authorship of code on a line level.

    \smallskip
    \noindent
    \item \textit{CommitInfluenceGraphMiner} applies PageRank to commits as proposed by Suzuki et al~\cite{pagerank-commits}. 
    \textit{CommitInfluenceGraphMiner} finds bug-fixing commits by searching for ``fix'' in commit messages~\cite{osman2014mining}. 
    Then, using \textit{Git blame}, the miner finds the earlier commits in which the lines changed in the current fix commit were also changed.  
    The output of the miner is a map of lists, with keys corresponding to fixing commit IDs and values, to the commits introducing the lines changed by the fixes.

    \smallskip 
    \noindent
    \item \textit{FileDependencyMatrixMiner} and \textit{AssignmentMatrixMiner} produce data for analysis of socio-technical congruence. 
    \textit{AssignmentMatrixMiner} is an auxiliary miner that yields a modification count for each \emph{developer + file} pair in the form of a nested map --- data that can be later used for calculation of socio-technical congruence~(\Cref{sec:congruence-description}).
    \textit{FileDependencyMatrixMiner} processes commits to find the files that were changed in the same commit. 
    For each pair of files, the miner yields a number of times they have been edited in the same commit.
    This data can be utilized to build the edges in the $G_s$ part of the socio-technical software network based on Conway's law (\Cref{sec:congruence-description},~\cite{Conway1967HOWDC}). 
    
    \smallskip 
    \noindent
    \item \textit{WorkTimeMiner} mines the distribution of commits over time in the week. 
    This data can be used, \eg to improve work scheduling by finding intersections in the time distributions between different developers. 
    
    \smallskip 
    \noindent
    \item \textit{ChangedFilesMiner} mines sets of changed files for each developer. It can be used, \eg for deriving sets of common edited files for multiple developers. 
    
\end{itemize}

\textbf{Calculation}. 
Some forms of data require non-trivial computations~\cite{carlson2015engaging}. 
To 
ensure extensibility, processing code is separated from the miners into dedicated classes. 
\miner features three calculation classes:
\begin{itemize}
    \item \textit{CoordinationNeedsMatrixCalculation} computes the coordination needs matrix according to the algorithm of~\cite{congruence}, using the data obtained by \textit{FileDependencyMatrixMiner} and \textit{AssignmentMatrixMiner}. 
    The computation results are represented as a matrix $C = \{C_{ij}\}$, where $i, j$ are the developer user IDs, and $C_{ij}$ is the relative coordination need~\cite{congruence}
    between the two individuals.
    \item \textit{MirrorCongruenceCalculation} computes the socio-technical congruence according to \cite{conway}, using the data yielded by \textit{CoordinationNeedsMatrixCalculation}. 
    Its output is a single number $C$, $0 \leq C \leq 1$, and higher values correspond to higher socio-technical congruence.
    \item \textit{PageRankCalculation} computes the PageRank vector according to the algorithm by Suzuki et al.~\cite{pagerank-commits}. A PageRank vector contains importance rankings 
    for each commit.
    The input data for \textit{PageRankCalculation} is the commit influence graph produced by \textit{CommitInfluenceGraphMiner}.
    The output is a vector where 
    each element represents the importance of a commit.
\end{itemize}

\textbf{Visualization}. 
\miner includes basic browser-based visualization generators for output of \textit{FileDependencyMatrixMiner} and \textit{CoordinationNeedsMatrixCalculation}.



\subsection{Extensibility}
Each class in \miner implements one of the interfaces described above.
This makes it easily extensible to cater for other mining tasks. 

\subsubsection{Additional miners}
A new miner can be implemented with minimal effort by extending the GitMiner abstract class. 

\begin{lstlisting}[
    language = Kotlin, 
    caption = An example implementation of a new GitMiner, 
    captionpos = b,
    label = lst:usageexample,
    basicstyle = \scriptsize]
    
class MyMiner(
    repository: FileRepository,
    neededBranches: Set<String>
) : GitMiner(repository, neededBranches, numThreads = 1) {
    var consecutiveCommits = 0
        private set

    override fun process(
        currCommit: RevCommit,
        prevCommit: RevCommit
    ) {
        val author1 = currCommit.authorIdent.emailAddress
        val author2 = prevCommit.authorIdent.emailAddress
        if (author1 == author2) consecutiveCommits++
    }
    
    override fun saveToJson(resourceDirectory: File) {...}
}
\end{lstlisting}


\subsubsection{Calculation}
To create a new calculation class, one can extend the \textit{Calculation} interface.
The interface contains of two functions: \textit{run} for executing calculations and \textit{saveToJson} for saving the result.

\subsubsection{Visualization}
To create a new visualisation class, one should extend the \textit{GraphHTML} interface. 

\subsection{Performance}
We tested TNM performance in mining lists of changed files per user against a trivial script based on pygit2\cite{pygit2} running on an average laptop. For a repository with $\approx53 000$ commits, mining with TNM took 42s compared to 6m 8s for the script. On a smaller repository with $\approx14 000$ commits mining with TNM and the script respectively took 3s and 25s. TNM outperformes pygit2 due to the multithreaded mining.




    
    
\section{Future work}
We are going to provide extensive documentation for \miner, further improve its capabilities of integration with arbitrary pipelines, and implement more processing features, such as automatic identity merging.


\bibliographystyle{IEEEtran}
\bibliography{tech-miner}

\end{document}